\documentclass[aps,prd,a4paper,eqsecnum,superscriptaddress,nofootinbib,showpacs,twocolumn,showkeys,amsfonts,amssymb,amsmath]{revtex4}

\usepackage{amssymb,latexsym}
\usepackage{amsmath, amsthm}

\begin{document}

\title{Thermodynamics and the naked singularity in the Gamma-metric}

\author{K. Lochan} \email{kinjalk@tifr.res.in}
\affiliation{Tata Institute of Fundamental Research, Homi Bhabha road, Colaba, Mumbai 400005, India}

\author{D. Malafarina} \email{daniele.malafarina@polimi.it}
\affiliation{Tata Institute of Fundamental Research, Homi Bhabha road, Colaba, Mumbai 400005, India}

\author{T. P. Singh} \email{tpsingh@tifr.res.in}
\affiliation{Tata Institute of Fundamental Research, Homi Bhabha road, Colaba, Mumbai 400005, India}


\begin{abstract}
\noindent We investigate a possible way of establishing a parallel between the third law of black hole mechanics, and the strong version of the third law of
 thermodynamics. We calculate the surface gravity and area for a naked singular null surface in the Gamma-metric and explain in what sense
this behaviour violates thermodynamics.
\end{abstract}


\maketitle

\section{Black Hole Thermodynamics and Cosmic Censorship}

\noindent We begin by making a few observations about the second and third law of black hole mechanics, and the third law of thermodynamics.

The proof of the second law of black hole mechanics [i.e. the area theorem] requires the validity of the cosmic censorship hypothesis.
 By now, various examples of the formation of naked singularities are known, in models of gravitational collapse studied in the general
theory of relativity. It is also known that general relativity admits extremal stationary Kerr-Newman black hole solutions, which posses in
fact naked singularities. What is the impact of these examples on the validity of the aforementioned proof of the second law? The general
 outlook in the community is that the Kerr-Newmann naked singularities are non-generic and hence they have no impact on the proof \cite{Wald2}.

It is well-known that the suggestive analogy between area and entropy becomes precise only when quantum effects [Hawking radiation]
are taken into account. Could it then not be the case that the area theorem itself strictly holds [in the generalized form where entropy
of the radiated matter is also taken into account] only when quantum effects are taken into account, the role of quantum effects now being
that of preventing the formation of naked singularities?

The third law of black hole mechanics was stated by Bardeen, Carter and Hawking [BCH] as: `It is impossible by any procedure,
 no matter how idealized, to reduce the surface gravity of a black hole to zero by a finite sequence of operations' \cite{bch}.
A precise formulation and proof of this law was given by Israel \cite{Israel}. The proof requires cosmic censorship to hold.
 The thermodynamic analog of this third law is the weak form of the third law of thermodynamics : `The temperature of a system
 cannot be reduced to zero in a finite number of operations'.

The strong form of the third law of thermodynamics states that: `As the temperature of a system goes to zero, its entropy goes to an
 absolute constant, which may be taken as zero'. The weak form of the third law follows as a consequence of the strong form. There is
no analog of the strong form of the third law of thermodynamics in black hole mechanics, because there are examples of extremal black
holes for which surface gravity goes to zero, but the area of the outer horizon remains non-zero and does not go to a universal constant
(recall that this is an example of a Kerr-Newmann naked singularity). It is also known that there are theoretical counterexamples to the
 strong form
of the third law in thermodynamics. One such example was constructed by Wald \cite{Wald} by considering a rotating bosonic gas at low
temperatures with a specific density of states. However, although some of these counterexamples can be regarded as `physically reasonable',
 they have not been experimentally realized.

What then should the analog between the third law of black hole mechanics and third law of thermodynamics be? There are various
 possible options to consider:

(i) The strong form of the third law of thermodynamics is not a law, because there are counterexamples to it. The analog should be between
 the weak form and the BCH statement of the third law of black hole mechanics, for which Israel has given a proof.

(ii) The counterexamples to the strong form of the third law of thermodynamics have not been realized experimentally, hence this form of
the law should be accepted only as empirically valid, notwithstanding the theoretical counterexamples. Furthermore, there possibly are reasons
 as to why extremal black-holes are `exotic objects' which are not continual limits of non-extremal black holes, and hence should not be
regarded as violations of a possible strong version of the third law of black hole mechanics [viz. as surface gravity goes to zero area
goes to an absolute constant]. This point of view has been nicely expressed by Racz \cite{Racz}.

(iii) The strong form of the third law of thermodynamics is a correct law, and there are (as yet unknown) reasons why the
theoretical counterexamples are not realized experimentally. Extremal black holes violate the strong version of the third law of black
hole mechanics, and they violate cosmic censorship. We propose that the strong version of the third law of black hole mechanics holds
when quantum effects are taken into account, the role of quantum effects being to avoid occurrence of naked singularities.

In summary, the observations that: (i) the analogy between area and entropy becomes precise only after quantum effects are taken into
 account, (ii) classical general relativity does admit some examples of formation of naked singularities in gravitational collapse,
(iii) formation of naked singularities is possibly avoided in quantum theory (quantum effects become important in the final stages of collapse)
 and (iv) the theoretical counterexamples to the third law of thermodynamics have not been realized empirically, lead us to suggest that:

The analogy between the strong version of the third law of thermodynamics and the strong version of the third law of black hole mechanics
does not hold at the theoretical level (because there are counterexamples) but it holds at the empirical level [counterexamples are not
realized in the real world].

 The idea that General Relativity can be thought as the theory of gravity that emerges at large scales from quantum statistical effects justifies the claim that the connection between thermodynamics and gravity must be valid in a more generic sense and not only for the event horizons. In this view the laws of thermodynamics first established for event horizons of stationary space-times may be generalized to `Isolated Horizons' \cite{Ashtekar}
and naked singularities might be investigated from a thermodynamical point of view.

On the other hand one could also say that naked singularities, which belong to another class of configurations, different from Black Holes
(be it because of initial data for gravitational collapse, or because of the values of the parameters in Kerr-Newman type spacetime),
prevent the precise connection between gravity and thermodynamics from being realized in a context more general than just black holes. In fact the laws of thermodynamics when extended to Killing/Isolated horizon of naked singularities seem to be violated or at least seem to contradict our experience.
However recent studies permit a thermodynamical interpretation of gravity \cite{thermogravity} and the association of an entropy with null surfaces \cite{null}.

In this note we present one such instance of a violation, by observing the properties of the Gamma-metric, which is a static axially
symmetric vacuum spacetime belonging to the Zipoy-Voorhees class \cite{Zipoy}. We point out that the metric has a
 singular null surface on which the surface gravity diverges, and which is also naked. Furthermore, the null surface has zero area.
Thus this surface in the Gamma-metric has zero entropy and a divergent surface gravity [equivalently temperature] - behaviour that is not in
accordance with what we expect from the laws of thermodynamics. One may hence conclude that the validity of thermodynamics, once assumed or demonstrated on a general level should prevent the possible occurrence of such naked singularities.

\section{The Gamma-metric}

\noindent The most general static, axially symmetric line element in cylindrical coordinates takes the form
\begin{equation}\label{weyl}
    ds^2=-e^{2\lambda}dt^2+e^{2\mu-2\lambda}[d\rho^2+dz^2]+r^2e^{-2\lambda}d\phi^2 \; ,
\end{equation}
where $\lambda(\rho, z)$ and $\mu(\rho, z)$ are solutions of
\begin{eqnarray}\label{laplace}
  \nabla^2\lambda &=& 0 \; ,\\
  \mu_{,\rho} &=& \rho(\lambda_{,\rho}^2+\lambda_{,z}^2) \; ,\\
  \mu_{,z} &=& 2\rho\lambda_{,\rho}\lambda_{,z} \; .
\end{eqnarray}
Here $\nabla$ is the flat Laplace operator in two dimensions implying that there is a one to one correspondence
between solutions of the Laplace equation in flat space and static axial symmetry; therefore in principle all static axially symmetric space-times are known. In fact, once we find $\lambda$ as a solution of equation \eqref{laplace}, then $\mu$ can be obtained by quadrature.\\

The Gamma-metric is the static axially symmetric space-time that corresponds to the solution of Laplace equation for a linear distribution
of matter of constant density $\frac{\gamma}{2}$ and length $2M$ distributed symmetrically along the $z$ axis. In Erez-Rosen coordinates
(or prolate spheroidal coordinates), given by $\rho^2 = (r^2-2Mr)\sin^2\theta$,  $z = (r-M)\cos\theta$ this solution takes the form:
\begin{equation}
    \lambda(r, \theta)=\frac{\gamma}{2}\ln\left(1-\frac{2M}{r}\right) \; ,
\end{equation}
to which corresponds
\begin{equation}
    \mu(r, \theta)=\frac{\gamma^2}{2}\ln\left(\frac{1-\frac{2M}{r}}{1-\frac{2M}{r}+\frac{M^2}{r^2}\sin^2\theta}\right) \; .
\end{equation}
The line element for the Gamma-metric can be written as:
\begin{eqnarray}
    ds^2&=&-\Delta^\gamma dt^2+\frac{\Delta^{\gamma^2-\gamma-1}}{\Sigma^{\gamma^2-1}}dr^2+ \\ \nonumber
    &&+r^2\Delta^{1-\gamma}
    \left[\left(\frac{\Delta}{\Sigma}\right)^{\gamma^2-1}d\theta^2+
    \sin^2\theta d\phi^2\right] \; ,
\end{eqnarray}
where
\begin{equation}
  \Delta = \left(1-\frac{2M}{r}\right), \; \; \Sigma = \left(1-\frac{2M}{r}+\frac{M^2}{r^2}\sin^2\theta\right) \, .
\end{equation}

The Gamma-metric reduces to the well known Schwarzschild space-time in spherical coordinates for $\gamma=1$ and represents the external field of
 an oblate (prolate) object for $\gamma>1$ ($\gamma<1$ respectively).


Furthermore the Gamma-metric exhibits a naked curvature singularity on the surface $r=2M$ when $\gamma\neq 1$ \cite{gamma}. Investigation of the Kretschmann scalar
shows that the singularity is a strong curvature singularity that can not be removed by a coordinate transformation. Furthermore the investigation of null geodesics terminating at the singular surface proves that the singularity has a directional behavior for $\gamma>1$. In fact the singularity in this case is not visible along the polar axis $\theta=0$, meaning that photons emitted by the singularity along this axis need an infinite time to reach far away observers.

\subsection{Isolated horizon}
An isolated horizon $\Xi$ is a closed, convex null surface which is in equilibrium i.e., across which there is no gravitational radiation or flux of matter fields. It is a 3-dimensional hypersurface in space-time with the following properties
\begin{itemize}
  \item[(i)] $\Xi$ is null, topologically $S^2 \times R$, and equipped with a preferred foliation by 2-surfaces $S_{\Xi}$ transverse to its null normal.
  \item[(ii)] All equations of motion are satisfied on $\Xi$.
  \item[(iii)] There is no flux of radiation either through or along $\Xi$.
               $$\mathring{R}_{ab}{\it l^b} = \mathring{R}_{ab} n^b =0,$$
             where $\mathring{R}_{ab}$ is pull back of $R_{ab}$ on  $S_{\Delta}$ and ${\it l^a},n^a$ are two null normals to  $S_{\Xi}$.
\item[(iv)]  $\Xi$ is non rotating, non-expanding future boundary of space-time under consideration.
\end{itemize}
In the case of the Gamma-metric the surface $ r= 2M$ (as a limit) satisfies these criteria. So, it is worth calculating thermodynamic quantities attached with this surface.

\subsection {Killing horizon in the Gamma-metric}

Being static and axially symmetric the Gamma-metric admits the two Killing vector fields $\partial_t$ and $\partial_\phi$.\\
The surface
\begin{equation}
    f=r-2M=0
\end{equation}
turns out to be a Killing horizon for $1+\gamma-\gamma^2 >0$ with the condition $\gamma > 0$, as shown below.\\
The fact that $f(r,\theta) = 0$ is a null surface follows from
\begin{equation}
 g^{\mu \nu}\bigtriangledown_{\mu}f\bigtriangledown_{\nu}f=0 \; ,
\end{equation}
which, once written for the Gamma-metric, becomes
\begin{equation}
 r^2\Delta^{\gamma-\gamma^2}
\Sigma^{\gamma^2-1}(\partial_\theta f)^2-\Delta^{\gamma-\gamma^2+1}\Sigma^{\gamma^2-1}(\partial_r f)^2=0 \; ,
\end{equation}
showing that $f=r-2M$ is in fact a null surface.

Furthermore it's easy to check that the Killing field $\partial_t$ is normal to $f$.
In fact if we write $\partial_t$ as $\xi^{\bf a} = \xi^{\mu} e_{\mu} ^{\bf a}$, then $\xi^{\mu} = \delta^{\mu}_t$ and therefore on $r=2M$ we get
\begin{equation}
\xi^{\bf a}\xi_{\bf a}|_S = g_{\mu \nu}\delta^{\mu}_t\delta^{\nu}_t =0 \; .
\end{equation}

We next proceed to compute the surface gravity for this Killing Horizon in the usual manner. The surface gravity is defined as
\begin{equation}
\kappa =\sqrt{-\frac{1}{2}(\bigtriangledown_{\mu}\xi_{\nu})(\bigtriangledown^{\mu}\xi^{\nu})} \; .
\end{equation}
For the Killing field $\partial_t$, the index lowered components are $\xi_{\mu} =g_{\mu \nu}\xi^{\nu} = g_{\mu t}$.
Since the Gamma-metric is diagonal, only $\xi_t$ survives and hence the only non vanishing component of $\xi_{\nu,\mu}$ is $\xi_{t,r}$.
\newline
It is easy to check that
\begin{equation}
\bigtriangledown_{t}\xi_{r} =- \Gamma^{t}_{r t}\xi_{t} = -\bigtriangledown_{t}\xi_{r} \; ,
\end{equation}
and
\begin{equation}
\bigtriangledown^{r}\xi^{t} =\frac{1}{2}g^{tt}g^{rr}\frac{\partial g_{tt}}{\partial r} \; .
\end{equation}
Hence,
\begin{eqnarray}
\kappa = \sqrt{-\frac{1}{4}g^{tt}g^{rr}\left(\frac{\partial g_{tt}}{\partial r}\right)^2} \; .
\end{eqnarray}
So that finally the surface gravity for the Gamma-metric has the form
\begin{equation}
\kappa =\frac{2M\gamma}{r^2}\Delta^{-\frac{(\gamma-1)^2}{2}}\Sigma^{\frac{\gamma^2-1}{2}} \; .
\end{equation}
Now the exponent of first term is negative for all values of $\gamma$, including those also for which $r=2M$ is a Killing Horizon.
Therefore, on this surface $\kappa $ diverges everywhere on the surface, except for $\theta =0,\pi$ possibly.

\section{Entropy of Killing Horizon}

For quite some time there have been attempts to model gravity as a thermodynamic theory \cite{thermogravity}, \cite{Paddy}. It has been suggested that any null surface acts as a horizon for a specialized set of observers, which traces out the degrees of freedom associated with the inaccessible region of spacetime.

Therefore, there is a notion of entropy associated with the null surfaces and  if gravity is an emergent thermodynamic property, it is demanded
that laws of thermodynamics should be applicable to any surface where Einstein's equations hold, specifically on all local Rindler causal horizons
through each spacetime point. Thus all Rindler horizons can be associated with a relation of the form $\delta Q = T\delta S$ where surface gravity and the area play the role of temperature and entropy respectively \cite{thermogravity}, \cite{null}.
In this sense Einstein's equations can be regarded as equations of state. Although this line of thinking calls into the attempts of quantization
of Einstein's equations it also suggests that, if true, some of the `configurations' of spacetime are thermodynamically not favourable.
In accordance with this line of thinking we can approach the $r=2M$ surface in a limiting fashion to test its `thermodynamic' properties.\\

We now explain how the area of the null surface under consideration has the interpretation of entropy.
It has been shown \cite{null} that by maximizing the entropy functional
\begin{equation}
 S = \int_{V} d^Dx \sqrt{-g}(4 P^{cd}_{ab}\bigtriangledown_c \xi^a \bigtriangledown_d \xi^b - T_{ab} \xi^a \xi^b)
\end{equation}
w.r.t. all the null vector fields $ \xi^a $ Einstein field equations are obtained as an `equilibrium configuration',
 where $ P^{cd}_{ab}$ is a fourth rank tensor built out
of the metric and other geometric
quantities which can be expanded in powers of number of derivatives of the metric
\begin{equation}
P_{abcd}(g_{ij},R_{ijkl})= c_1 P_{abcd}^{(1)}(g_{ij}) +c_2 P_{abcd}^{(2)}(g_{ij},R_{ijkl})+ . . .
\end{equation}
with the $m$-th order $P_{a}^{(m) ijk}$ related to $m$-th order Lagrangian density of Lanczos-Lovelock gravity
\begin{equation}
P_{a}^{(m) ijk}= \frac{\partial \mathcal{L}_{m}^{(D)}}{\partial R^{a}_{ijk}} \; ,
\end{equation}
with
\begin{equation}
\mathcal{L}_{m}^{(D)} = \frac{1}{16\pi} 2^{-m} \delta^{a_1a_2..a_{2m}}_{b_1b_2..b_{2m}}R^{b_1b_2}_{a_1a_2}...R^{b_{2m-1}b_{2m}}_{a_{2m-1}a_{2m}} \; .
\end{equation}

Choosing $c_m = 0$, $\forall m\neq 1$,  and $c_1 =1$ Einstein field equations can be recovered.
The tensor $P^{cd}_{ab}$ is analogous to the elastic constant in case of elastic solids and has the same symmetry properties as the Riemann tensor.
Also the following equation is true at each order:
\begin{equation}
\bigtriangledown_d P^{cd}_{ab} =0 \; .
\end{equation}

The entropy of the Killing horizon is calculated in the Rindler limit. $S\arrowvert_{On\;Shell}$ is calculated for a particular (non-null) surface in
the Rindler
frame and then this surface is made to approach the Killing horizon; thus the entropy of the horizon is obtained in a limiting sense.
 By $S\arrowvert_{On\;Shell}$
we mean that equilibrium configuration is assumed, which for this case is the field equations (without correction terms from higher order
 $P_{ab}^{(m) cd}$).
For this purpose
$\xi^a$ is chosen normal to the surface and
\begin{equation}
S\arrowvert_{\Xi} =4\pi\int_{\Xi} d^2 x_{\perp}\sqrt{g^{(2)}}\mathcal{L}^{(2)}_{0} \; .
\end{equation}
In the case of the Gamma-metric we have a naked curvature singularity at $ r= 2M $. Hence a Rindler frame cannot be established at the surface.

Therefore, one should calculate the entropy in the limiting case only.
Now $\mathcal{L}^{(2)}_{0} = \frac{1}{16 \pi}$, hence the entropy of null surfaces as we approach the Killing horizon is
\begin{equation}
 S\arrowvert_{\Xi} \longrightarrow \frac{1}{4}\int_{\Xi} d^2 x_{\perp}\sqrt{g^{(2)}} \; .
\end{equation}
which is proportional to (in fact $\frac{1}{4}$th of) the area of the $r=2M $ surface in the Gamma-metric.

\subsection{Area of revolution surfaces in the Gamma-metric}

The area of a spheroid of a fixed radius $r>2M$ in the Gamma space-time is
\begin{equation}
    A_{\gamma}=\int_0^{2\pi}\int_0^\pi r^2\sin\theta\Delta^{\frac{(1-\gamma)^2}{2}}
    \Sigma^{\frac{1-\gamma^2}{2}}d\theta d\phi \; .
\end{equation}
For $\gamma=1$ it reduces to the area of the Schwarzschild spheres
\begin{equation}
    A_S=4\pi r^2 \; .
\end{equation}
When $\gamma\neq 1$ we see that the reduced metric on the 2-surface $r=$const. and $t=$ const. becomes degenerate on the singular surface $r=2M$
 and therefore
\begin{equation}
    A_\gamma|\xrightarrow[r\longrightarrow2M]{} 0 \; .
\end{equation}
 Since the area goes to zero, the entropy tends to zero too.
Since the surface gravity diverges on this surface, we may conclude that the combination of `infinite temperature and zero entropy'
is not consistent with usual thermodynamics, and hence the occurrence of such a singularity might be forbidden by thermodynamics.

\section{Conclusions}
\noindent In this brief note we have suggested that the analogy between the strong form of the third law of thermodynamics, and the third law of black-hole mechanics, might be taken seriously since counterexamples to them do not seem to be realized in the real world.
We have also suggested that the applicability of the third law of thermodynamics may extend to gravitational configurations other than black holes as well, as seen above in the example of the naked singularity in the Gamma-metric. Here, the surface gravity [analog of the temperature] diverges on the Killing horizon, whereas the area [analog of the entropy] goes to zero. Consistency with the third law would require surface gravity to go to zero when the area goes to zero. Since that does not happen here, this example of a naked singularity could be regarded as possessing an
anti-thermodynamic feature, and hence being unrealistic. Similarly it is reasonable to suppose that this behaviour is due to the choice of the class of space-times to which the Gamma-metric belongs (namely vacuum, static and axially symmetric) and therefore it will be shared by other examples of manifolds belonging to the same class. If that is the case then the anti-thermodynamic behaviour, i.e. the fact that this kind of naked singularities may not emerge as a thermodynamic limit from the quantum theory of gravity, could well be regarded as a generic feature of space-times with naked singularities.
We note though that strictly speaking what we have here is a self-consistency argument, and not a deduction, since a thermodynamic interpretation of gravity itself indeed requires cosmic censorship to hold [we recall the importance of Rindler horizons in the development of this interpretation].

It will be interesting to examine how naked singularities that arise in dynamical gravitational collapse [such as the shell-focusing naked singularity in the Lemaitre-Tolman-Bondi spherical dust collapse] relate to the third law of thermodynamics. Can one construct analogs of surface gravity and area in this context, and examine whether the strong form of the third law is being obeyed or not? This issue is at present under investigation.

\end{document}